\documentclass{jpp}
\usepackage{graphicx}

\usepackage[utf8]{inputenc}
\usepackage[T1]{fontenc}
\usepackage{amsmath}

\usepackage[colorlinks=true]{hyperref}

\DeclareMathOperator{\sign}{sgn}

\shorttitle{Effect of insulator end cap thickness on Hartmann flow}
\shortauthor{R. Gaur, I. G. Abel, B. Tripathi, and E. Kolemen}

\title{Effect of insulator end cap thickness on time-dependent Hartmann flow in a rotating mirror}

\author{Rahul Gaur\aff{1}
  \corresp{\email{rg6256@princeton.edu, rgaur@terpmail.umd.edu}},
  Ian G. Abel\aff{2},
  Bindesh Tripathi\aff{3},
  \and Egemen Kolemen\aff{1,4}}

\affiliation{\aff{1} Department of Mechanical and Aerospace Engineering, Princeton University, Princeton, USA
\aff{2} Institute for Research in Electronics and Applied Physics, University of Maryland, College Park, USA 
\aff{3} Department of Physics, University of Wisconsin, Madison, USA
\aff{4} Princeton Plasma Physics Laboratory, Princeton, USA}

\begin{document}

\maketitle

\begin{abstract}
   We present a framework for analyzing plasma flow in a rotating mirror. By making a series of physical assumptions, we reduce the magnetohydrodynamic (MHD) equations in a three-dimensional cylindrical system to a one-dimensional system in a shallow, cuboidal channel within a transverse magnetic field, similar to the Hartmann flow in the ducts. We then solve the system both numerically and analytically for a range of values of the Hartmann number and calculate the dependence of the plasma flow speed on the thickness of the insulating end cap.  We observe that the mean flow overshoots and decelerates before achieving a steady-state value, a phenomenon that the analytical model cannot capture. This overshoot is directly proportional to the thickness of the insulating end cap and the external electric field, with a weak dependence on the external magnetic field. Our simplified model can act as a benchmark for future simulations of the supersonic mirror device Compact Magnetic Fusion Experiment (CMFX), which will employ more sophisticated physics and realistic magnetic field geometries. 
\end{abstract}

\section{Introduction}
Rotating mirrors present a novel research direction in magnetic confinement fusion~\citep{post1987magnetic}. This approach typically employs a large background magnetic field combined with an externally supplied current, generating plasma rotation at supersonic speeds. Compared with more complex magnetic configurations, this method offers a potentially simpler design for fusion reactors. Previous experiments have demonstrated the feasibility of supersonic rotating mirrors for nuclear fusion, as reported in the literature~\citep{reid2014100, ellis2005steady}.

The CMFX (Compact Magnetic Fusion Experiment), funded by the ARPA-E BETHE (Breakthroughs Enabling THermonuclear-fusion Energy) program~\citep{romero2021overview}, aims to achieve a significant fusion triple product --- with the product of plasma density, temperature, and confinement time greater than $10^{20} keV s/m^3$. The success of this endeavor would mark a substantial advance in the development of practical fusion energy.

The effectiveness of this approach may depend on the choice of materials for the plasma end caps. These disk-shaped components are critical for maintaining axial plasma confinement and enabling magnetic field rotation. Therefore, the material characteristics of the end caps are important for the efficiency and viability of the fusion process.

In this study, we examine how the electrical properties and thickness of the end-cap materials affect plasma flow. To simplify the analysis, we assume a model that transforms the problem from laminar flow in a cylindrical geometry to laminar flow in a shallow, cuboidal channel within a transverse magnetic field, similar to Hartmann flow in ducts~\citep{hartmann1937hg, hartmann1937hg2}.

\begin{figure}
    \centering
    \includegraphics[width=0.45\linewidth, trim = 0 0mm 0 0mm, clip]{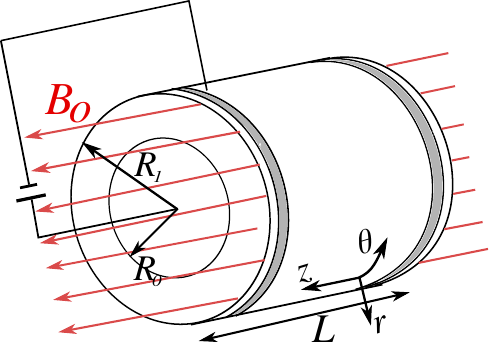} 
    \caption{We illustrate a simplified version of the supersonic mirror in a cylindrical coordinate system $(r, \theta, z)$. The background field $B_0 \hat{\boldsymbol{z}}$ is generated by external magnets (not shown). The device comprises an inner electrode, which is a solid conducting rod of radius $R_0$, and an outer conducting shell of radius $R_1$. On both ends, the gray part of the end cap denotes an insulator, whereas the white part denotes an imperfect conductor. Plasma remains in the annular region between the two electrodes. Due to the external potential difference between the electrodes, the radial current $j_{\mathrm{ext}} \hat{\boldsymbol{r}}$ flows through the plasma; coupled with the background field $\boldsymbol{B}_0 = B_0 \hat{\boldsymbol{z}}$ causes the plasma to rotate in the azimuthal ($ \widehat{\boldsymbol{\theta}}$) direction.}
    \label{fig:CMFX}
\end{figure}

Previous research, including the work by~\citet{huang2004magnetohydrodynamic}, has studied the impact of the Hartmann boundary layer on plasma flow and magnetohydrodynamic (MHD) stability.~\citet{hassam2019multiscale} have also explored the effect of perfectly conducting walls and the emergence of small-scale physical oscillations in a one-dimensional MHD channel flow. Recent work by~\citet{zhang2022magnetohydrodynamic} has analyzed flow regimes in rectangular channels over a wide range of magnetic and fluid Reynolds numbers. However, a comprehensive solution that encompasses end-cap fields and their impact on flow remains elusive. Our simplified model aims to establish an analytical relationship, providing a benchmark for future simulations employing more sophisticated physics and realistic magnetic field geometries. The principles derived here are adaptable to broader scenarios using complex numerical or analytical models. Figure~\ref{fig:CMFX} illustrates the setup of the supersonic mirror device.

To describe the evolution of the flow under an external azimuthal force, we use the MHD model. In Section~\ref{sec:section-2}, we explain the three-dimensional MHD model and the assumptions used to simplify it to a one-dimensional model. We then solve the one-dimensional model in all three media: imperfect conductor, insulator, and plasma. We present the evolution of the time-dependent solution at different values of the Hartmann number. In Section~\ref{sec:section-3}, we demonstrate how the mean plasma flow overshoots and decelerates, before reaching a steady-state value, and how the thickness of the insulator and the external fields affect the mean flow. In Section~\ref{sec:section-4}, we present our conclusions.

\section{Solving the MHD model: Theory and analysis}
\label{sec:section-2}
In this section, we present a theoretical framework for analyzing plasma behavior in a rotating mirror. We adopt the incompressible MHD model, which assumes a constant plasma density, to describe plasma dynamics. To facilitate our analysis, we initially simplify the model to a one-dimensional (1D) representation based on a set of well-defined assumptions appropriate to the CMFX device. The resulting simplified 1D model is then solved numerically, and further simplified and solved analytically to provide insights into the plasma behavior under the influence of the rotating magnetic field in the mirror setup.

For plasma, we start with the incompressible MHD~\citep{IdealMHD} model
\begin{equation}
    \boldsymbol{\nabla}\cdot \boldsymbol{u} = 0,
    \label{eqn:ideal-MHD-continuity}
\end{equation}
\begin{equation}
    \rho \left(\frac{\partial \boldsymbol{u}}{\partial t} + \boldsymbol{u}\cdot \boldsymbol{\nabla} \boldsymbol{u}\right) = -\boldsymbol{\nabla} p + \boldsymbol{j} \times \boldsymbol{B} + \mu \boldsymbol{\nabla}^2 \boldsymbol{u} + \boldsymbol{F},
    \label{eqn:ideal-MHD-momentum}
\end{equation}
\begin{equation}
    \left( \frac{\partial}{\partial t} + \boldsymbol{u}\cdot \boldsymbol{\nabla} \right) \frac{p}{\rho^{\gamma}}  = 0, 
    \label{eqn:ideal-MHD-pressure-evolution}
\end{equation}
\begin{equation}
    -\frac{\partial \boldsymbol{B}}{\partial t} = \boldsymbol{\nabla} \times \boldsymbol{E}, 
    \label{eqn:ideal-MHD-induction}
\end{equation}
\begin{equation}
    \mu_{\rm{p}} \boldsymbol{j} = \boldsymbol{\nabla} \times \boldsymbol{B} - \frac{1}{c^2}\frac{\partial \boldsymbol{E}}{\partial t}, 
    \label{eqn:MHD-Ampere}
\end{equation}
\begin{equation}
    \boldsymbol{E} = -\boldsymbol{u}\times \boldsymbol{B} +\eta \boldsymbol{j},
    \label{eqn:Ohm's-law}
\end{equation}
\begin{equation}
    \boldsymbol{\nabla}\cdot \boldsymbol{B} = 0,
    \label{eqn:div-B}
\end{equation}
where $\rho$ is the plasma density, $\boldsymbol{u}$ is the plasma flow velocity, $\boldsymbol{B} = \boldsymbol{B}_0 + \boldsymbol{B}_1$ is the total magnetic field, $\boldsymbol{B}_0$ is the static background field generated by the external coils, $\boldsymbol{B}_1$ is the time-dependent field generated by the plasma, $\boldsymbol{j}$ is plasma-generated current density, $p$ is the plasma pressure, $\boldsymbol{F} = \boldsymbol{j}_{\mathrm{ext}} \times \boldsymbol{B}_0$ is the external force on the plasma generated by the current $\boldsymbol{j}_{\mathrm{ext}}$ generated due to the externally applied potential difference. The constants $\mu$, $\eta$, and $\mu_{\rm{p}}$ are the dynamic viscosity, magnetic field diffusivity, and magnetic permeability of the plasma, respectively. In equilibrium, we assume a uniformly magnetized static plasma with $p = p_0, \rho = \rho_0, \boldsymbol{u} = 0, \boldsymbol{B} = \boldsymbol{B}_0$. 

Similarly, inside the plasma wall materials described in Figure~\ref{fig:CMFX}, we solve the Maxwell's equations
\begin{equation}
    -\frac{\partial \boldsymbol{B}}{\partial t} = \boldsymbol{\nabla} \times \boldsymbol{E},
\end{equation}
\begin{equation}
    \mu_{\mathrm{c}} \boldsymbol{j} = \boldsymbol{\nabla} \times \boldsymbol{B} - \frac{1}{c^2}\frac{\partial \boldsymbol{E}}{\partial t},
    \label{eqn:Amperes-law-conductor}
\end{equation}
where $\mu_{\mathrm{c}}$ is the magnetic field permeability of the imperfect conductor. By definition, inside an insulator, the current
\begin{equation}
    \boldsymbol{j} = 0,
    \label{eqn:Amperes-law-insulator}
\end{equation}
and for an imperfect conductor that satisfies Ohm's law, 
\begin{equation}
    \boldsymbol{j} = \sigma_{\mathrm{c}} \boldsymbol{E},
    \label{eqn:Ohms-law-conductor}
\end{equation}
where $\sigma_{\mathrm{c}}$ denotes the conductivity of the imperfect conductor. The last two equations are the result of the electrical properties of the end-cap materials.
The electric and magnetic fields in all three regions: imperfect conductor, insulator, and plasma must be continuous, since there is no charge accumulation or surface current on the interfaces. We begin by normalizing the model equations and describing the various assumptions we make to simplify the three-dimensional MHD model.

\subsection{Normalization and cylinder-to-slab transformation}
In this section, we simplify the model by normalizing it as follows
\begin{equation}
    \widetilde{\boldsymbol{B}} = \frac{\boldsymbol{B}}{B_0}, \quad \widetilde{\boldsymbol{u}} = \frac{\boldsymbol{u}}{v_{\mathrm{A}}}, \quad v_{\mathrm{A}} = \frac{B_0}{\sqrt{\mu_{\mathrm{p}} \rho_0}}, \quad \widetilde{\boldsymbol{E}} = \frac{\boldsymbol{E}}{v_{\mathrm{A}} B_0}, \quad \tilde{t} = \frac{t v_{\mathrm{A}}}{L}, \quad  \tilde{z} = \frac{z}{L}, \quad \tilde{r} = \frac{r}{R_0},
    \label{eqn:normalization-1}
\end{equation}
where $B_0$ is the magnetic field due to the coils, $v_A$ is the Alfv\'{e}n speed, $L$ is axial length, and $R_0$ is the radial size of the device. 
Solving~\eqref{eqn:ideal-MHD-continuity}-\eqref{eqn:Ohms-law-conductor} analytically in a three-dimensional cylindrical geometry is not possible. To reduce the complexity of this model, we adopt an asymptotic ordering that differentiates between multiple scales and various quantities by employing the small parameter $\epsilon$
\begin{equation}
    \tilde{u}_r, \tilde{u}_z \sim  \epsilon \tilde{u}_{\theta} \sim  \epsilon^2, 
    \quad  \tilde{B}_r, \tilde{B}_z \sim \epsilon \tilde{B}_{\theta} \sim \epsilon \tilde{B}_{\mathrm{v}0} \sim  \epsilon^2, \quad \tilde{E}_{r} \sim \epsilon \tilde{E}_{\mathrm{v}0} \sim \epsilon^2, 
    \label{eqn:orderings-1}
\end{equation}
\begin{equation}
    \quad  \widehat{\boldsymbol{z}}\cdot\boldsymbol{\nabla} \sim 1/L \sim (\widehat{\boldsymbol{r}}, \widehat{\boldsymbol{\theta}})\cdot \boldsymbol{\nabla} \sim 1/R_0, \quad  (R_1-R_0)/R_0 \sim \epsilon,\quad
    \beta \sim \epsilon^2,\quad  \lvert \boldsymbol{F} \rvert \sim \epsilon.
    \label{eqn:orderings-2}
\end{equation}

In these orderings, the subscript letter $r, \theta$, or $z$ denotes the component of a quantity, while the subscript $\mathrm{v}$ denotes the external fields in the vacuum region, and $\beta = 2 \mu_o p/B_0^2$ is the ratio of plasma pressure to magnetic pressure. The assumptions $\hat{r}\cdot \boldsymbol{\nabla}\sim 1/R$ and $(R_1-R_0)/R_0 \sim \epsilon$ implies a shallow device with small spatial gradient that enables us to eliminate any radial variation in our annular domain. By making assumptions that remove radial variation and impose azimuthal symmetry, the problem in three dimensions can be effectively transformed into a configuration that resembles a one-dimensional slab or rectangular channel.

We also assume that the azimuthal flow dominates the other components of the flows and that the plasma-generated azimuthal magnetic field is greater than the rest of the components of the field. Such behavior of the plasma leads to a set of equations where all the nonlinear terms involving interaction between the flow and fields completely vanish, transforming our model into a set of linear equations. Another important assumption to be used is that the power supply generates electric and magnetic fields $E_{\mathrm{v}}$ and $B_{\mathrm{v}}$ that are of the same order as the plasma-generated fields. These fields will be an important part of our conclusion in the last section. 

\begin{figure}
    \centering
    \begin{tabular}[b]{c} 
        \includegraphics[width=0.45\linewidth, trim = 0 0mm 0 0mm, clip]{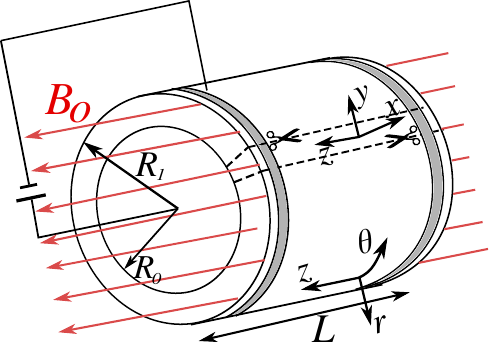} \\
        {\textit{(a)} Rotating mirror setup}
    \end{tabular} 
  \hspace*{-3mm}
  \begin{tabular}[b]{c}
        \includegraphics[width=0.27\linewidth, trim = 0 0mm 0 0mm, clip]{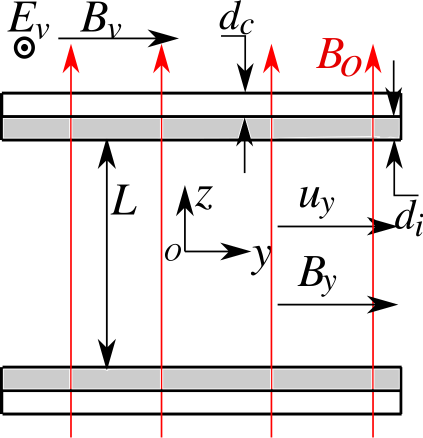} \\[2mm]
        {\textit{(b)} A simplified slab (rectangular channel) model}
  \end{tabular} \\[1mm]
    \caption{In figure $\textit{(a)}$, we illustrate the setup of a large aspect ratio mirror device in the form of an annular cylindrer. In figure $\textit{(b)}$, we present a cut section of the mirror in figure $\textit{(a)}$, representing a simplified slab system. The region outside the end caps is treated as a vacuum with external vacuum fields $E_{\mathrm{v}}, B_{\mathrm{v}}$. The equilibrium field $\boldsymbol{B}_0$ points in the $z$-direction whereas the plasma flow $u_y$ and plasma-generated magnetic field $B_y$ are in the azimuthal $y$-direction.}
    \label{fig:Hartmann_flow}
\end{figure}

The conceptualization and simplification of the three-dimensional domain is further elucidated in Figure~\ref{fig:Hartmann_flow}.
In the slab limit, we substitute the subscripts $r, \theta$, and $z$ with $x, y$, and $z$, respectively. Note that such equations only involve the radial component of the electric field $\tilde{E}_r(z)$(or $\tilde{E}_x(z)$ in a slab), the azimuthal components of the magnetic field $\tilde{B}_{\theta}(z)$(or $\tilde{B}_y(z)$ in a slab) and plasma flow $\tilde{u}_{\theta}(z)$(or $\tilde{u}_y(z)$ in a slab) throughout this paper.
To further simplify the notation, we omit $\, \widetilde{\phantom{}}\, $ in normalized quantities.
In the following section, we begin by solving the Maxwell's equation in an imperfectly conducting end cap.

Throughout the paper, we impose even parity on the flow and electric field and odd parity on the magnetic field about the $z=0$ plane,
\begin{equation}
\begin{gathered}
 u_y(z) = u_y(-z), \quad E_x(z) = E_x(-z), \quad B_y(z) = -B_y(-z).
\end{gathered}
\label{eqn:parity-conditions}
\end{equation}

\subsection{Time-dependent fields in the vacuum region}
We start by assuming the the fields outside the device  
\begin{equation}
\begin{gathered}
    E_{\mathrm{v}} = E_{\mathrm{v}0} [1 - \exp(-c_0 t)],\\
    B_{\mathrm{v}} = B_{\mathrm{v}0} \sign(z) [1 - \exp(-c_0 t)],\\
\end{gathered}
\label{eqn:E_v-B_v}
\end{equation}
where $B_{\mathrm{v}0}, E_{\mathrm{v}0}$ are the steady-state vacuum electric and magnetic fields due to external sources such as the power supply, we can use the outside fields as boundary conditions to calculate the coefficients in the rest of media.

The external electric and magnetic fields arise from the voltage difference and the current flowing through the power lines. The time-dependent form of the fields is similar to that of the current and potential of a resistor circuit. The parameter $c_0$ corresponds to the time taken for a typical experimental discharge, typically hundreds of Alfv\'en time periods. 

These fields will be used as boundary conditions for the time-dependent fields inside the imperfect conductor in the next section.

\subsection{Time-dependent solution inside the imperfect conductor}
\label{subsec:imperfect-conductor-solution}
Our analysis begins by solving the Maxwell's equations within the imperfect conductor. We employ the magnetic and electric fields in the vacuum region as boundary conditions to fully determine the field characteristics up to the interface between the imperfect conductor and the insulator. Using the induction equation
\begin{equation}
    \frac{\partial B_{y}}{\partial t} = -\frac{\partial E_x}{\partial z},
    \label{eqn:induction-equation-imperfect-conductor}
\end{equation}
and the Ampere's law
\begin{equation}
    E_x = \frac{1}{\mathrm{S}_{\mathrm{c}}} j_{x} = -\frac{1}{\mathrm{S}_{\mathrm{c}}}\partial_z B_{y}, \quad \mathrm{S}_{\mathrm{c}} = v_{A} L \sigma_{\mathrm{c}} \mu_{\mathrm{c}},
\end{equation}
where $\mu_{\mathrm{c}}$ is the magnetic field permeability of the imperfect conductor, $\sigma_{\mathrm{c}}$ is the electric conductivity, and $\mathrm{S}_{\mathrm{c}}$ is the Lundquist number which is a measure of magnetic diffusion inside a material. We have neglected the displacement current term because it scales as $(v_{\mathrm{A}}/c)^2 \ll 1$, which is small compared on the Alfv\'{e}nic time scale $t \sim 1$ of the problem.  

The general fields inside the imperfect conductor can be written as
\begin{equation}
\begin{gathered}
    E_x = -\frac{D_3}{\mathrm{S}_{\mathrm{c}}} + \sqrt{\frac{c_0}{\mathrm{S}_{\mathrm{c}}}} \exp(-c_0 t)\left[  D_5 \cos(\sqrt{\mathrm{S}_{\mathrm{c}} c_0}z) + D_6 \sign(z) \sin(\sqrt{\mathrm{S}_{\mathrm{c}} c_0}z) \right],\\
    B_y = D_3 z + D_4 - \exp(-c_0 t)\left[ D_5 \sin(\sqrt{\mathrm{S}_{\mathrm{c}} c_0} z) - D_6 \sign(z) \cos(\sqrt{\mathrm{S}_{\mathrm{c}} c_0} z) \right].
\end{gathered}
\label{eqn:imperfect-conductor-1}
\end{equation}
Matching the tangential fields $E_x$ and $B_y$ with the outside fields $E_{\mathrm{v}}$ and $B_{\mathrm{v}}$, respectively, we obtain and simplify the fields inside the conductor
\begin{equation}
\begin{split}
    E_x = E_{\mathrm{v}0} - \exp(-c_0 t)& \bigg[E_{\mathrm{v}0} \cos(\sqrt{\mathrm{S}_{\mathrm{c}}c_0}(z-\sign(z)(0.5+d_{\mathrm{i}}+d_{\mathrm{c}}))) \\ 
    &  +\sqrt{\frac{c_0}{\mathrm{S}_{\mathrm{c}}}} B_{\mathrm{v}0} \sign(z) \sin(\sqrt{\mathrm{S}_{\mathrm{c}}c_0}(z-\sign(z)(0.5+d_{\mathrm{i}}+d_{\mathrm{c}})))\bigg],
\end{split}
\label{eqn:imperfect-conductor-E_x}
\end{equation}
\begin{equation}
\begin{split}
    B_y =  -\mathrm{S}_{\mathrm{c}} E_{\mathrm{v}0} &(z - \sign(z)(0.5 +d_{\mathrm{i}} +d_{\mathrm{c}})) + B_{\mathrm{v}0} \sign(z)\, + \\
    \exp(-c_0 t)&\bigg[\sqrt{\frac{\mathrm{S}_{\mathrm{c}}}{c_0}}E_{\mathrm{v}0}\sin(\sqrt{\mathrm{S}_{\mathrm{c}}c_0}(z-\sign(z)(0.5+d_{\mathrm{i}}+d_{\mathrm{c}}))) \\
    &- B_{\mathrm{v}0} \sign(z) \cos(\sqrt{\mathrm{S}_{\mathrm{c}}c_0}(z-\sign(z)(0.5+d_{\mathrm{i}}+d_{\mathrm{c}})))\bigg].
\end{split}
\label{eqn:imperfect-conductor-B_y}
\end{equation}
Equations~\eqref{eqn:imperfect-conductor-E_x} and~\eqref{eqn:imperfect-conductor-B_y} represent the electromagnetic fields within the imperfect conductor. We note that the expressions include a $\sign$ function that is discontinuous at $z = 0$. However, since the domain of the imperfect conductor does not include the point $z = 0$, the fields remain continuous inside the conductor, and the use of a $\sign$ function is valid. Note that these solutions are only valid for finite values of $S_{\mathrm{c}}$. These results will serve as boundary conditions for the subsequent analysis of the fields within the insulator, as explained in the following section.
\subsection{Time-dependent solution inside the insulator}
\label{subsec:insulator-solution}
After solving for the field inside the imperfect conductor, we solve for the electromagnetic fields inside the insulator end cap. The fields inside the perfect insulator must satisfy the induction equation
\begin{equation}
    \frac{\partial B_{y}}{\partial t} = -\frac{\partial E_x}{\partial z},
\end{equation}
and the Ampere's law
\begin{equation}
    j_{x} = -\partial_z B_{y} = 0.
\end{equation}
In the insulator, the magnetic field is spatially uniform, whereas the electric field varies in response to the time-dependent magnetic field. The tangential components of the magnetic $B_{y}$ and electric $E_{x}$ fields
\begin{equation}
\begin{split}
    E_{x} &= E_{\mathrm{v}0} + \exp(-c_0 t) \Bigg\{\left[-E_{\mathrm{v}0} \cos(\sqrt{\mathrm{S}_{\mathrm{c}} c_0}d_{\mathrm{c}}) + \sqrt{\frac{c_0}{\mathrm{S}_{\mathrm{c}}}} B_{\mathrm{v}0} \sin(\sqrt{\mathrm{S}_{\mathrm{c}} c_0}d_{\mathrm{c}})  \right]\\
    &- c_0 \left(\lvert z \rvert - 0.5 - d_{\mathrm{i}}\right)\left[\sqrt{\frac{\mathrm{S}_{\mathrm{c}}}{c_0}} E_{\mathrm{v}0} \sin(\sqrt{\mathrm{S}_{\mathrm{c}} c_0}\, d_{\mathrm{c}}) + B_{\mathrm{v}0} \cos(\sqrt{\mathrm{S}_{\mathrm{c}} c_0}\, d_{\mathrm{c}}) \right] \Bigg\}.
    \label{eqn:E_y_insulator}
\end{split}
\end{equation}
\begin{equation}
\begin{split}
    B_y &= \sign(z)(\mathrm{S}_{\mathrm{c}} E_{\mathrm{v}0}\, d_{\mathrm{c}} + B_{\mathrm{v}0}) \\
    & - \sign(z) \exp(-c_0 t)\left[\sqrt{\frac{\mathrm{S}_{\mathrm{c}}}{c_0}} E_{\mathrm{v}0} \sin(\sqrt{\mathrm{S}_{\mathrm{c}} c_0}\, d_{\mathrm{c}}) + B_{\mathrm{v}0} \cos(\sqrt{\mathrm{S}_{\mathrm{c}} c_0}\, d_{\mathrm{c}}) \right].
    \label{eqn:B_y_insulator}
\end{split}
\end{equation}
The magnetic field stays the same as the boundary between the imperfect conductor and the insulator but the electric field changes linearly with the insulator thickness due to the induction equation. 

Similar to the previous section, we have neglected the displacement current term because it scales as $(v_{\mathrm{A}}/c)^2 \ll 1$, which is small compared on the Alfv\'{e}nic time scale $t \sim 1$ of the problem. However, it is important to note that close to $t=0$, the fields $E_{\mathrm{v}} = B_{\mathrm{v}} = 0$, whereas the fields inside the imperfect conductor and insulator have finite values which violates causality. This violation of causality arises from the omission of the displacement current term. In Appendix~\ref{app:Causality}, we show that incorporating the displacement current resolves this issue, and neglecting the displacement current does not influence the long-time solution or any outcomes of this study.

The effect of these external fields on the plasma dynamics is explored in Section~\ref{sec:section-3}.

\subsection{Time-dependent solution inside the plasma}
\label{subsec:plasma-solution}
After solving the fields in the imperfect conductor and the insulator, we solve the equation inside the plasma. This will completely define the solution in all three media.

The lowest order equations correspond to the equilibrium: $p = p_0, \rho = \rho_0, \boldsymbol{B} = \boldsymbol{B}_0,$ and $\boldsymbol{u} = 0$, and time-dependent quantities arise only at first order. The plasma satisfies~\eqref{eqn:ideal-MHD-continuity}-\eqref{eqn:div-B} which, using the orderings defined in~\eqref{eqn:orderings-1}-\eqref{eqn:orderings-2} for a shallow-channel case with dominant azimuthal plasma flow and magnetic fields, can be reduced to a set of coupled one-dimensional partial differential equations 
\begin{equation}
    \partial_t u_{y} = \partial_z B_{y}+ \frac{1}{\rm{Re}} \partial^2_{z} u_{y} + F_0(1 - \exp(-c_0 t)), \quad \mathrm{Re} = \frac{\rho_0 v_{\mathrm{A}} L}{\mu},
    \label{eqn:ideal-MHD-momentum1}
\end{equation}
\begin{equation}
    \partial_t B_{y} = \partial_{z} u_{y} + \frac{1}{\rm{Rm}} \partial^2_{z}B_{y}, \quad \mathrm{Rm} = \frac{v_{\mathrm{A}} L \mu_{\mathrm{p}}}{\eta},
    \label{eqn:ideal-MHD-induction2}
\end{equation}
\begin{equation}E_{x} = -u_{y} - \frac{1}{\rm{Rm}} \partial_{z} B_{y}, 
\label{eqn:Ohms-law-1}
\end{equation} 
subject to time-dependent boundary conditions on the insulator-plasma interface. The flow $u_{y} = u_{y}(z)$, and field $B_{y} = B_{y}(z)$, and the external forcing term $F_y = F_0 (1 - \exp(-c_0 t))$. Note that the form of the forcing function is chosen to be similar to that of the start-up stage of a rotating mirror. The forcing and external fields $E_{\mathrm{v}}, B_{\mathrm{v}}$ are generated by discharging a voltage source, typically an array of capacitors and therefore have the same time evolution, \textit{i.e.}, $(1-\exp(-c_0t))$. For simplicity, we assume that the electrical properties of the plasma do not change during operation. 

Coupled equations~\eqref{eqn:ideal-MHD-momentum1} and~\eqref{eqn:ideal-MHD-induction2} describe the evolution of the azimuthal plasma flow and magnetic field, while~\eqref{eqn:Ohms-law-1} determines the electric field. We also introduce a new dimensionless parameter, the Hartmann number, defined as $\mathrm{Ha} \equiv \sqrt{\mathrm{Re}\mathrm{Rm}}$, which will be used in the subsequent analysis. To further understand this model, we numerically solve it and provide analysis of the solution in the next section.
\section{Results from the simplified 1D Hartmann model}
\label{sec:section-3}
In this section, we solve the simplified one-dimensional Hartmann flow model and present a detailed analysis of our results. We will then explain how plasma flow depends on the insulator wall thickness.

\begin{figure}
    \centering
    \begin{tabular}[b]{c} 
        \includegraphics[width=0.43\linewidth, trim = 0 0mm 0 0mm, clip]{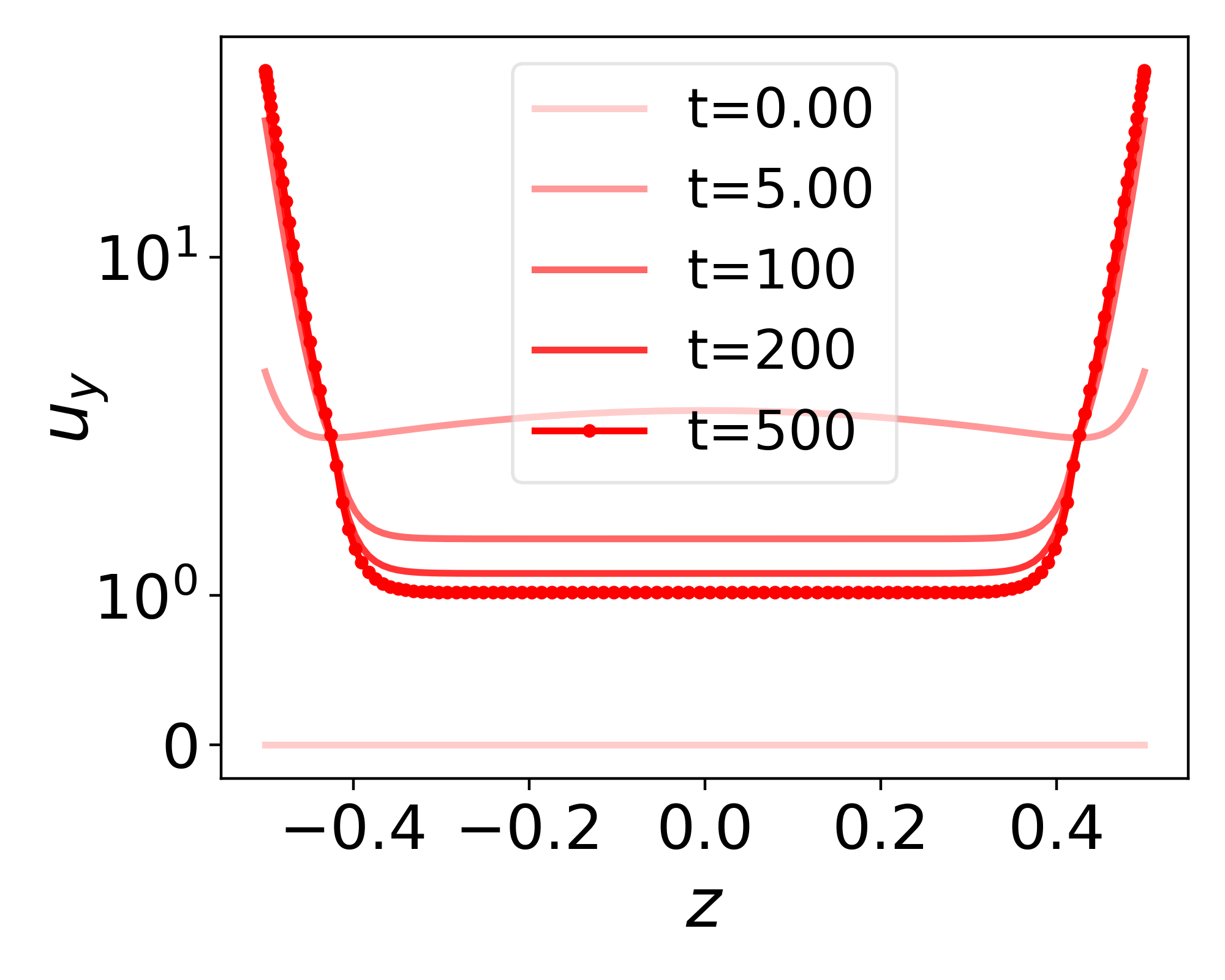} \\
    \end{tabular} 
  \hspace*{-4mm}
  \begin{tabular}[b]{c}
        \includegraphics[width=0.43\linewidth, trim = 0 0mm 0 0mm, clip]{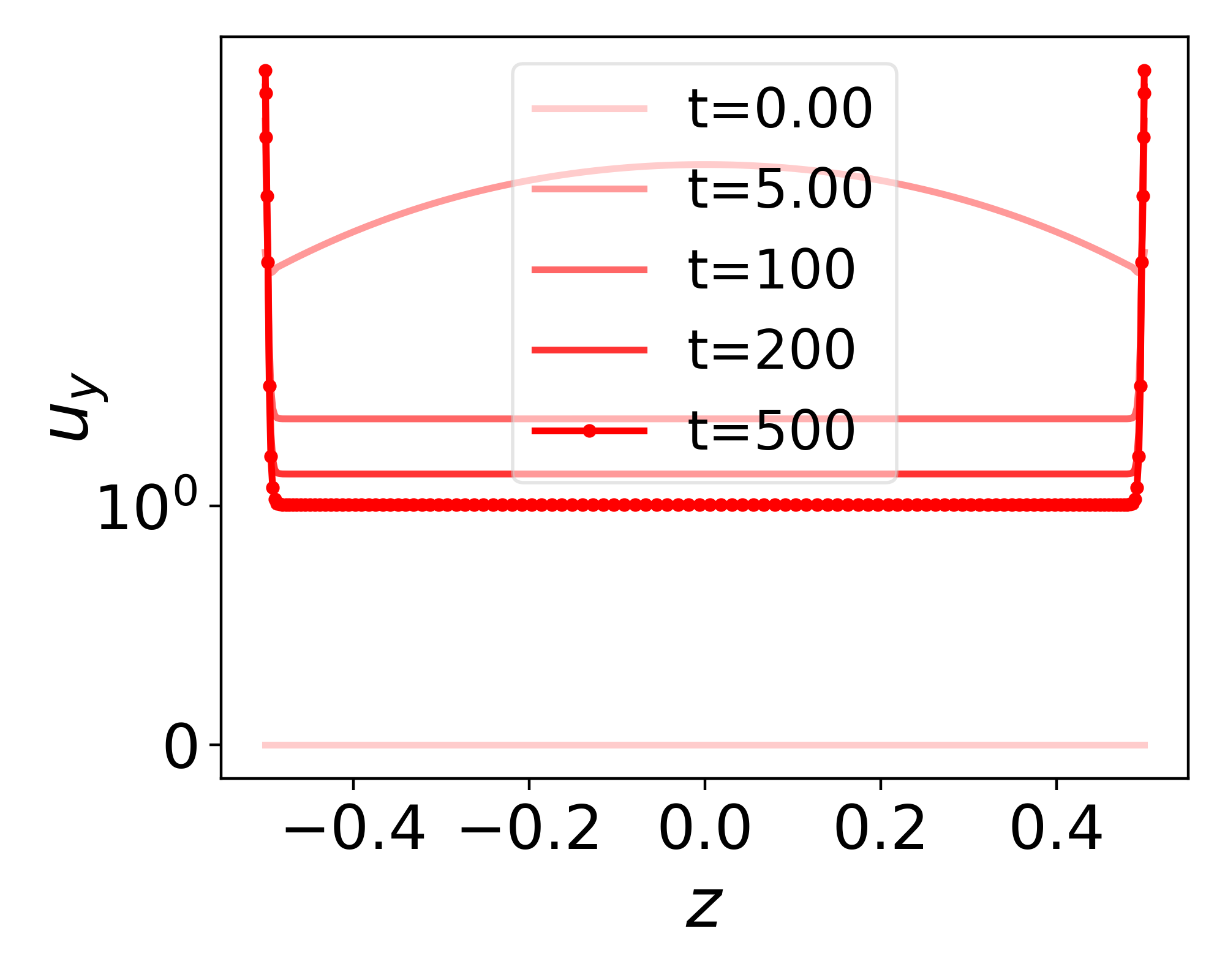} \\
  \end{tabular} \\[-3mm]
  \hspace*{-3mm}
    \begin{tabular}[b]{c} 
        \includegraphics[width=0.43\linewidth, trim = 0 0mm 0 0mm, clip]{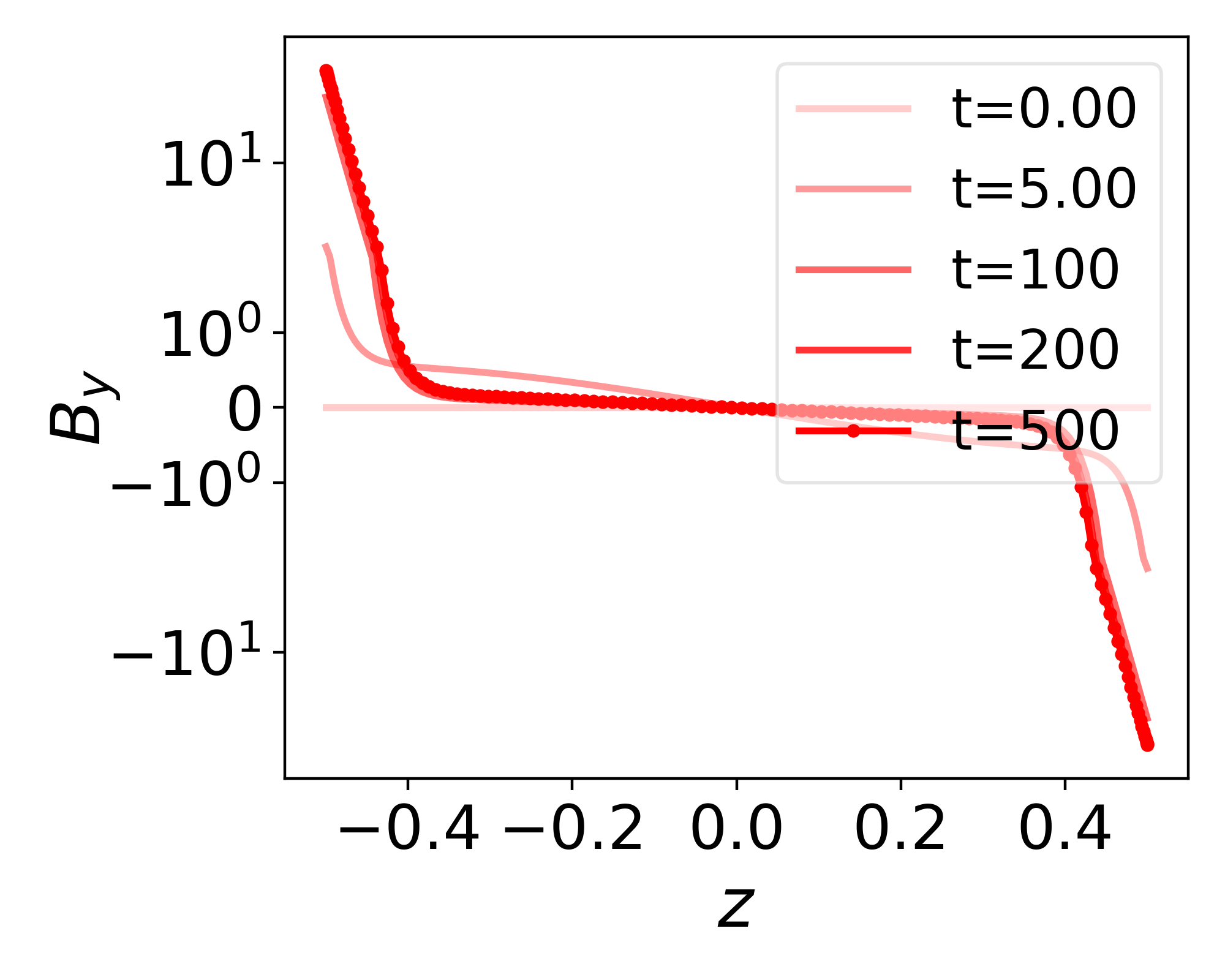} \\[-3mm]
        {\textit{(a)} $\mathrm{Re} = 50,  \mathrm{Ha} =  50$}
    \end{tabular} 
  \hspace*{-4mm}
  \begin{tabular}[b]{c}
        \includegraphics[width=0.43\linewidth, trim = 0 0mm 0 0mm, clip]{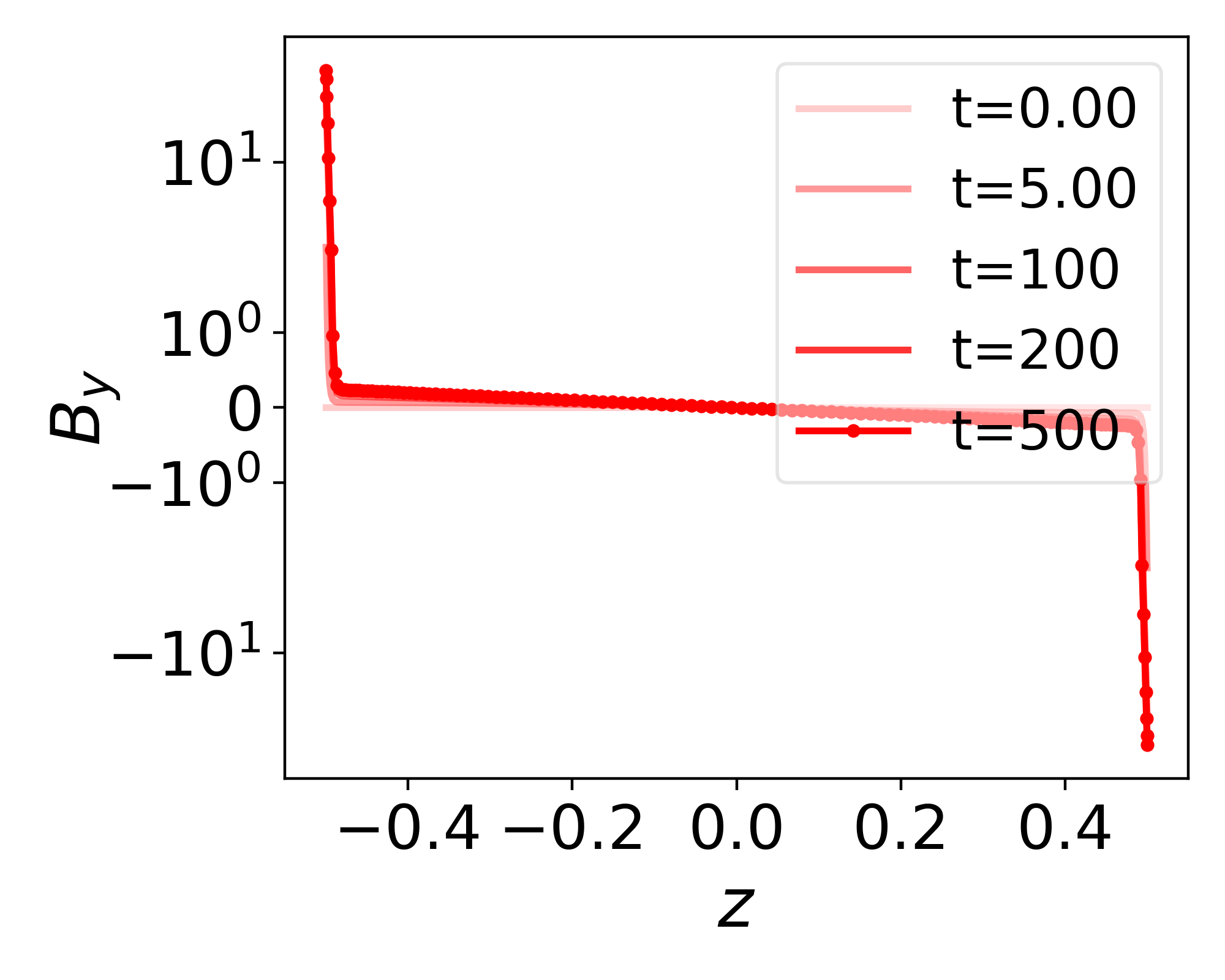} \\[-3mm]
        {\textit{(b)} $\mathrm{Re}= 50,  \mathrm{Ha} = 500$}
  \end{tabular}
    \caption{We plot the plasma flow and magnetic field profiles as functions of cylindrical distance $z$ at three different time values with $\textit{(a)}$ $\mathrm{Ha} = 50$ and $\textit{(b)}$ $\mathrm{Ha} = 500$. For this solution, we have chosen $F_0 = 0.5, c_0 = 10^{-2}, \mathrm{S}_{\mathrm{c}} = 10^{4}, d_{\mathrm{c}} = 5 \times 10^{-3}, d_{\mathrm{i}} = 10^{-2}, E_{\mathrm{v}0} = -1, B_{\mathrm{v}0} = 1$. Due to non-ideal effects, the plasma forms a sharp boundary layer near the insulating end caps. Note that the fields and flows have been scaled by $1/\epsilon$ to avoid adding factors of $\epsilon$ to all quantities on the $y$ axis.}  
    \label{fig:uy_and_By_z}
\end{figure}

We numerically solve~\eqref{eqn:ideal-MHD-momentum1}-\eqref{eqn:ideal-MHD-induction2} using the spectral numerical code Dedalus~\citep{burns2020dedalus} implementing time-dependent boundary conditions on electric and magnetic fields obtained from the solution in the insulating end caps, given by equations~\eqref{eqn:B_y_insulator} and~\eqref{eqn:E_y_insulator} at $z = \pm 0.5$. We apply a no-slip boundary condition to the flow. We use a Chebyshev grid with $n_z = 128$ grid points to resolve rapidly varying spatial features, such as the boundary layer. The Chebyshev polynomial is well suited for this problem, as the grid points are mostly clustered at the edge of the domain, which efficiently resolves the boundary layer. For the time-stepper, we use a second-order backward difference (SBDF2) time-stepping routine. An implicit method such as SBDF2 helps us avoid the unphysical oscillations associated with a stiff system such as this. We solve the model for two values of the Hartmann number and present the results in Figure~\ref{fig:uy_and_By_z}.

The plasma behavior is governed by ideal MHD in the core, with non-ideal effects dominating the boundary layer. The forcing term is balanced by the curvature of the magnetic field in the core region (around $z=0$), which is frozen in the fluid, flowing with a nearly uniform speed. However, in the boundary layer, the plasma can move relative to the magnetic field lines, which allows it to slip over the boundary. The thickness of the boundary layer is proportional to $1/\sqrt{\mathrm{Ha}}$ --- a system with a high $\mathrm{Ha}$ has a thinner boundary layer. Additionally, the core flow speed appears to be unaffected by the Hartmann number.

Since $\mathrm{Ha} \gg 1$ for supersonic rotating plasmas, the presence of a boundary layer introduces a new length scale that we can use to solve the problem analytically~\citep{bender2013advanced}. This process, the analytical solution, and its comparison with the numerical solution are presented in Appendix~\ref{app:Analytical_solution}.

\begin{figure}
    \centering
    \begin{tabular}[b]{c} 
        \includegraphics[width=0.42\linewidth, trim = 2mm 2mm 2mm 2mm, clip]{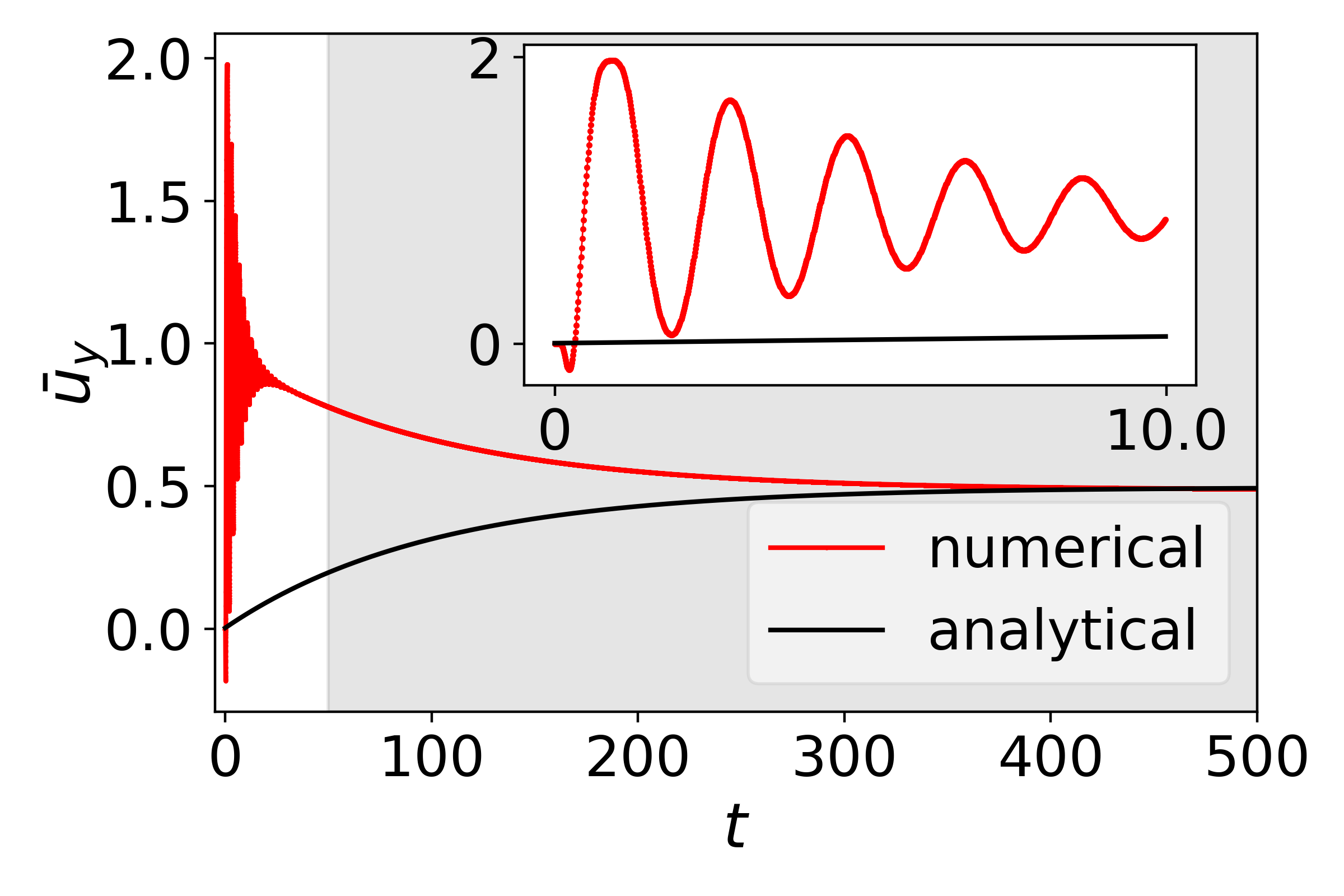} \\
        {\textit{(a)} $\mathrm{Ha} = 50$}
    \end{tabular} 
  \begin{tabular}[b]{c}
        \includegraphics[width=0.42\linewidth, trim = 2mm 2mm 2mm 2mm, clip]{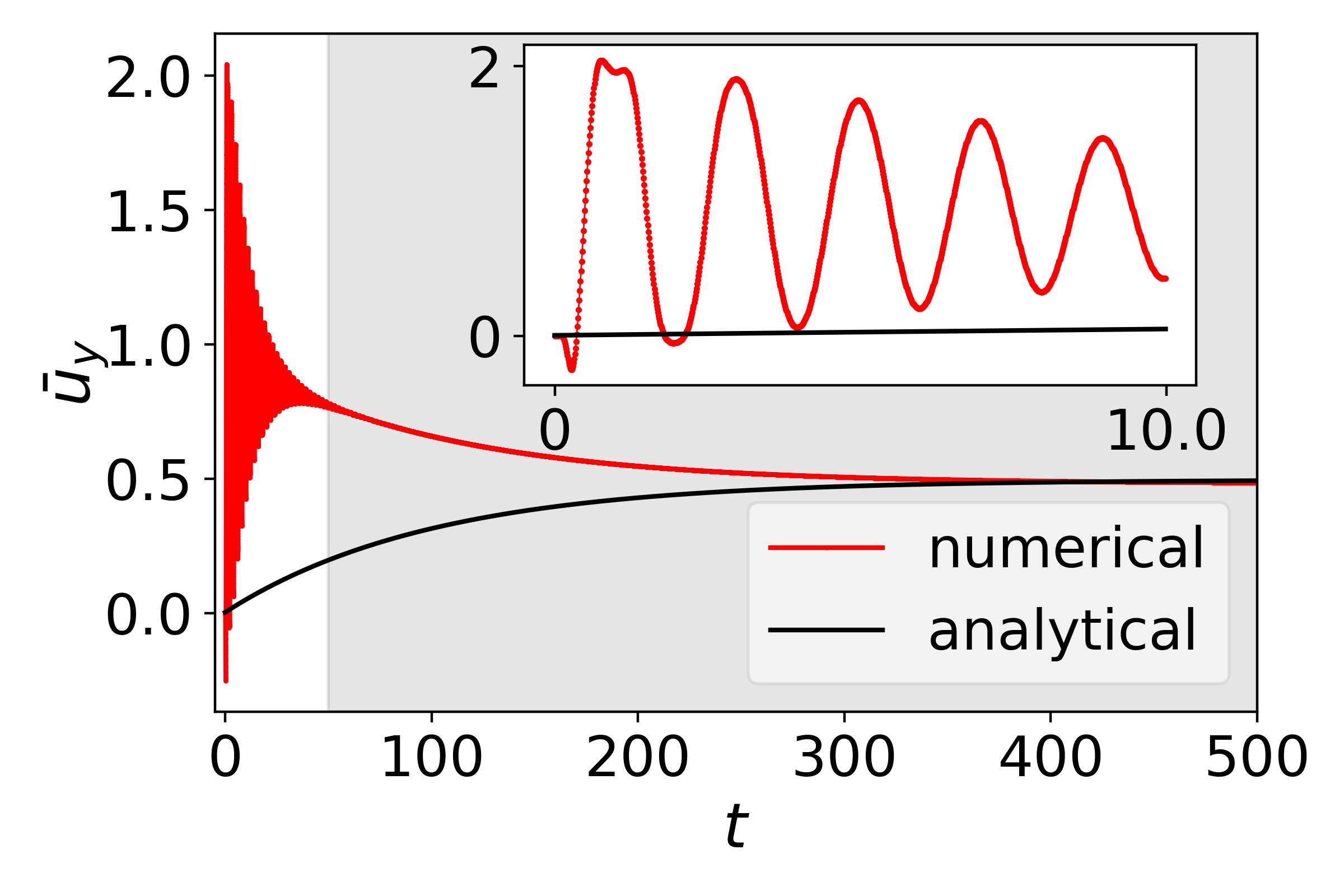} \\
        {\textit{(b)} $\mathrm{Ha} = 500$}
  \end{tabular}
    \caption{In this figure, we plot the mean core plasma flow speed $\bar{u}_y = 2\int_{-0.25}^{0.25} dz\, u_y $ as a function of time $t$ for in Figure~$\textit{(a)}$ for $\mathrm{Ha} = 50$ and in Figure~$\textit{(b)}$ for $\mathrm{Ha} = 500$ and compare the numerical and the simplified analytical solutions. The inset shows the initial part of the numerical and analytical solutions. The solutions agree well, but only close to the steady state. However, the analytical model cannot capture the Alfv\'{e}nic dynamics in the beginning, or the overshooting and subsequent deceleration (shaded region) of the flow. The parameters used for these figures are the same as the ones used in Figure~\ref{fig:uy_and_By_z}.}  
  \label{fig:acceleration}
\end{figure}

An important feature that we observe is the overshooting of the mean core plasma flow 
\begin{equation}
 \bar{u}_y = 2 \int_{-0.25}^{0.25} dz\, u_y, 
\end{equation}
beyond its steady state value. We demonstrate this phenomenon in Figure~\ref{fig:acceleration}.
As the boundary conditions are time-dependent, the system adjusts to the new boundary values through the generation of Alfv\'{e}n waves. However, time-dependent boundary conditions lead to an overshooting of the mean flow before it decelerates to a steady-state value. We observe this phenomenon in our model over a wide range of input parameters, and overshooting continues to occur in systems with high Hartmann numbers, regardless of whether the mean flow is calculated over the entire domain or just the core. Therefore, for the CMFX device, it is crucial to ensure a sub-Alfvénic mean flow, as approaching Alfv\'{e}nic speeds can induce various instabilities and reduce the confinement time~\citep{teodorescu2010sub}.

\begin{figure}
    \centering
    \begin{tabular}[b]{c}
        \includegraphics[width=0.4\linewidth, trim = 0 0mm 0 0mm, clip]{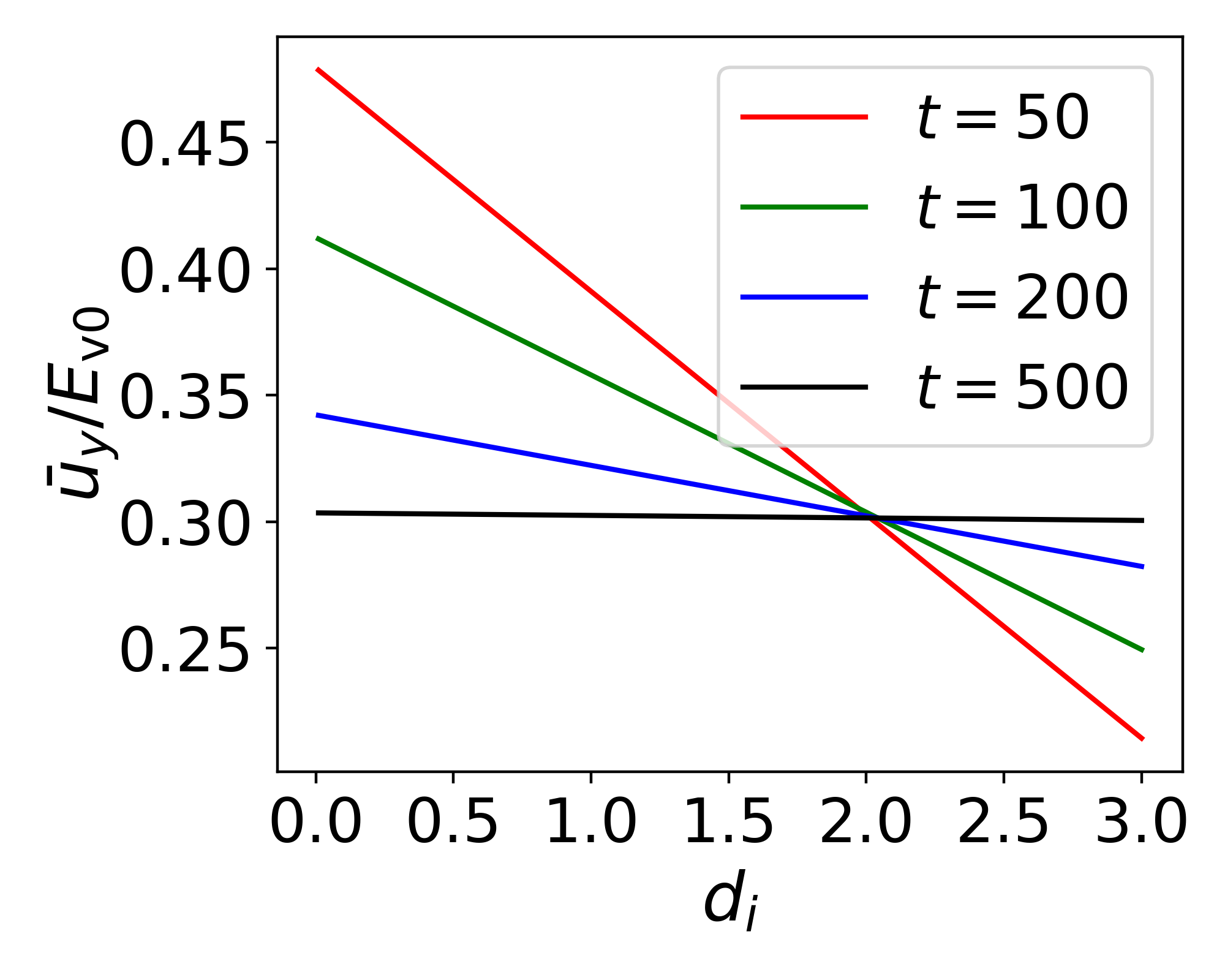} \\
        {\textit{(a)} $\mathrm{Ha} = 500, B_{\mathrm{v}0} = 1$}
    \end{tabular} 
    \begin{tabular}[b]{c} 
        \includegraphics[width=0.4\linewidth, trim = 0 0mm 0 0mm, clip]{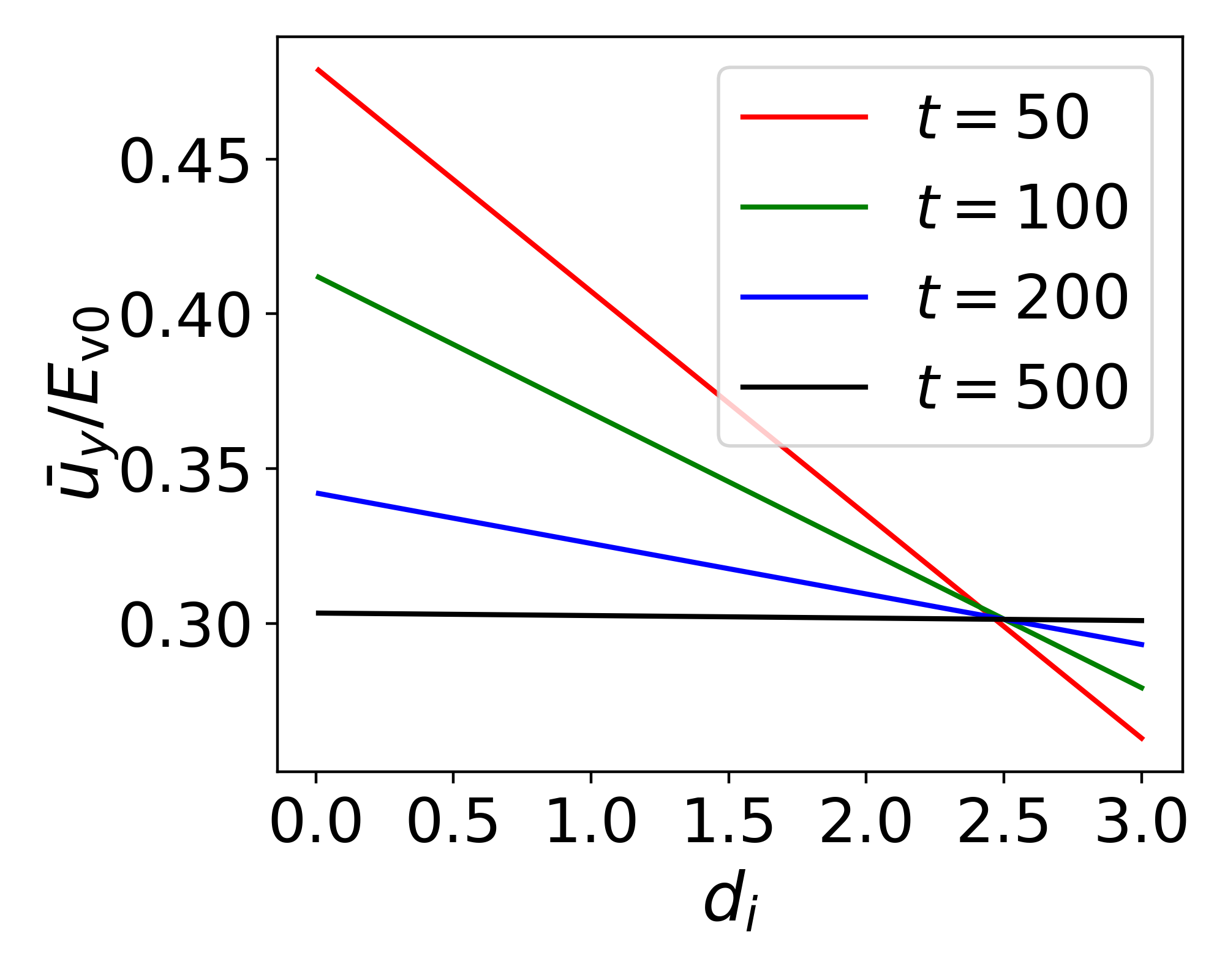} \\
        {\textit{(b)} $\mathrm{Ha} = 500, B_{\mathrm{v}0}  = 10$}
    \end{tabular} \\
    \caption{In this figure, we plot the dependence of the mean flow $\bar{u}_y$ normalized by the vacuum electric field $E_{\mathrm{v}0}$ against the insulator end cap thickness $d_i$ at different times and for different values of the external magnetic field. The flow velocity has a strong dependence on the thickness of the insulator at the beginning. We see that the flow transitions from overshooting to undershooting the steady-state value with increasing insulator thickness, around $d_i = 2$ in $5\textit{(a)}$ and $d_i = 2.5$ for $5\textit{(b)}$. Hence, to avoid overshooting the flow velocity, one must choose a thicker insulator end cap. Note that all plasma-generated flows and fields have been scaled up by $1/\epsilon$ because of their small size compared to the respective background quantities.}  
    \label{fig:uy_di}
\end{figure}

Finally, for the same parameter values used in Figure~\ref{fig:uy_and_By_z}, we also plot the dependence of the mean flow normalized by the external electric field $\bar{u}_y/E_{\mathrm{v}0}$ as a function of the thickness of the insulator $d_{\mathrm{i}}$ in Figure~\ref{fig:uy_di} for two different values of the external magnetic field $B_{\mathrm{v}0}$. 
We find that the magnitude of the plasma flow reduces with increasing insulator end cap thickness and is sensitive to the values of the external electric and magnetic fields, especially during the ramp-up phase of the device. However, the dependence of the mean flow on the insulator thickness is strongly affected by the external vacuum electric field $E_{\mathrm{v}0}$. Therefore, to avoid overshooting of the plasma flow from the target value, one must carefully choose the insulator end cap thickness. As the system approaches steady state, the dependence of the mean flow on the insulator goes down and in steady state, they are completely independent. 

To accurately determine the flow magnitude, we must accurately and self-consistently calculate the fields $E_{\mathrm{v}0}$ and $B_{\mathrm{v}0}$. However, in general, the electric field $E_{\mathrm{v}0} = E_{\mathrm{v}0}(V_0, R_0, R_1)$ is a complex function of $V_0$, the potential difference across the electrodes, and $R_0$, $R_1$, the radial positions of the electrodes. The current $I_{\mathrm{v}0} = I_{\mathrm{v}0}(V_0, \sigma_{\mathrm{p}}, \mu_{\mathrm{p}})$ in the external wires is an unknown function of $V_0$ and the electrical properties of the plasma. The magnetic field between the circular electrodes will be $B_{\mathrm{v}0} = B_{\mathrm{v}0}(I_{\mathrm{v}0}, R_0, R_1, \sigma_{\mathrm{p}}, \mu_{\mathrm{p}})$. Hence, the values of the fields depend on the electrical properties of the plasma and how they change over time. To better understand it, one would require a nonlinear, multi-physics solver.

\section{Conclusions}
\label{sec:section-4}
We simplified the three-dimensional MHD model by transforming the equations describing a supersonic rotating plasma in cylindrical geometry into a one-dimensional slab model by making a set of assumptions appropriate for the CMFX. We then numerically and analytically solved our simplified model and calculated the dependence of the plasma flow speed on the thickness of the insulating end cap. Our analysis shows that the mean flow speed of the plasma is linearly reduced with the thickness of the insulator wall. This reduction in mean flow is stronger for large external electric fields. Moreover, to avoid the plasma flow from overshooting beyond its steady-state value, one must carefully choose the insulator end cap thickness. Therefore, the design of the CMFX device must carefully take into account the electrical properties of the end caps and the external electric and magnetic fields.

This work opens many avenues for future research. A key step forward is to extend our technique to calculate the effect of the insulator thickness on the plasma flow by solving the 3D equations in all the different materials simultaneously using a multiphysics solver. Our simplified model can act as a benchmarking tool for these sophisticated solvers. The analysis could also be repeated with different boundary conditions, realistic magnetic field geometry, with temperature dependence, gyroviscosity, Hall effects, and kinetic effects.

Furthermore, our analysis has potential implications for ongoing liquid lithium-metal experiments, as detailed in~\citet{saenz2022divertorlets}, which are crucial for liquid divertor concepts in future tokamak fusion power plants. The similarity of these experiments with our one-dimensional slab model solution further underscores the relevance of our findings in practical fusion applications.

\textbf{Acknowledgments}: One of the authors, R.G., enjoyed fruitful discussions with Dr. Wrick Sengupta, Dr. Yi-Min Huang, Tony Qian, and Prof. Prateek Gupta. This research used the computing resources provided by the Stellar cluster at Princeton University.

\textbf{Declaration of interests}
The authors report that they do not have a conflict of interest.

\textbf{Data availability statement} {The Python scripts used to generate the results presented in this paper are freely available at \href{https://doi.org/10.5281/zenodo.11349455}{https://doi.org/10.5281/zenodo.11349455}}

\appendix
\section{Effect of ignoring displacement current at $t=0$}
\label{app:Causality}
As we are addressing the model on the time scale associated with the Alfv\'{e}n speed, we have neglected the faster time scale associated with the speed of light, which emerges from the displacement current term in Ampere's law. This results in an apparent violation of causality when, at $t = 0$, the fields in the vacuum region are zero, but the fields inside the end caps are finite. In this appendix, we show that including the displacement current effect and solving Maxwell's equations on a faster time scale resolves this issue.  

We assume the same normalizations and orderings as described in~\eqref{eqn:normalization-1},~\eqref{eqn:orderings-1}, and~\eqref{eqn:orderings-2} and include the displacement current term
in Ampere's law
\begin{equation}
    E_x = \frac{1}{\mathrm{S}_{\mathrm{c}}} j_{x} = -\frac{1}{S_{\mathrm{c}}}\partial_z B_{y} - \left(\frac{v_{\mathrm{A}}}{c}\right)^2 \frac{\partial E_x}{\partial t},
\end{equation}
along with the induction equation~\eqref{eqn:induction-equation-imperfect-conductor} , which gives us the following equation
\begin{equation}
    \left(\frac{v_{\mathrm{A}}}{c}\right)^2\frac{\partial^2 E_x}{\partial t^2} + \frac{\partial E_x}{\partial t} = \frac{1}{S_{\mathrm{c}}} \partial^2_z E_x.
\end{equation}
On a very short time scale $t \sim (v_{\mathrm{A}}/c)^2$, we solve
\begin{equation}
    \left(\frac{v_{\mathrm{A}}}{c}\right)^2\frac{\partial^2 E_x}{\partial t^2} + \frac{\partial E_x}{\partial t} = 0.
    \label{eqn:short-time-scale-equation}
\end{equation}
On the longer time scale $t \sim 1$, we solve
\begin{equation}
    \frac{\partial E_{x}}{\partial t} = -\frac{\partial^2 E_x}{\partial z^2},
\end{equation}
and use the solutions from Sections~\ref{subsec:imperfect-conductor-solution} and~\ref{subsec:insulator-solution}. Here we apply the boundary layer technique on two different time scales instead of spatial scales, as done in Appendix~\ref{app:Analytical_solution}. The general analytical solution to~\eqref{eqn:short-time-scale-equation} is $E_x = A (1- \exp(-t (c/v_{\mathrm{A}})^2))) E_x(x)$ where $E_x(x)$ is the spatial part that matches the long-time solution.
Using the separation of variables, the equations can be solved to obtain the following solution
\begin{equation}
\begin{split}
    E_x &= [1-\exp(-t (c/v_{\mathrm{A}})^2)]E_{\mathrm{v}0} \\
    &- \bigg\{\left[\exp(-c_0 t) - \exp(-t (c/v_{\mathrm{A}})^2) \right]\bigg[E_{\mathrm{v}0} \cos(\sqrt{\mathrm{S}_{\mathrm{c}}c_0}(z-\sign(z)(0.5+d_{\mathrm{i}}+d_{\mathrm{c}}))) \\ 
    &  +\sqrt{\frac{c_0}{\mathrm{S}_{\mathrm{c}}}} B_{\mathrm{v}0} \sign(z) \sin(\sqrt{\mathrm{S}_{\mathrm{c}}c_0}(z-\sign(z)(0.5+d_{\mathrm{i}}+d_{\mathrm{c}})))\bigg] \bigg\},
\end{split}
\end{equation}
\begin{equation}
\begin{split}
    B_y &= [1-\exp(-t (c/v_{\mathrm{A}})^2)]  \left[-\mathrm{S}_{\mathrm{c}} E_{\mathrm{v}0} (z - \sign(z)(0.5 + d_{\mathrm{i}} +d_{\mathrm{c}})) + B_{\mathrm{v}0} \sign(z)\, \right] \\
    &+ \bigg\{ \left[\exp(-c_0 t) - \exp(-t (c/v_{\mathrm{A}})^2)\right] \bigg[\sqrt{\frac{\mathrm{S}_{\mathrm{c}}}{c_0}}E_{\mathrm{v}0}\sin(\sqrt{\mathrm{S}_{\mathrm{c}}c_0}(z-\sign(z)(0.5+d_{\mathrm{i}}+d_{\mathrm{c}})))\, \\
    &\, \, - B_{\mathrm{v}0} \sign(z) \cos(\sqrt{\mathrm{S}_{\mathrm{c}}c_0}(z-\sign(z)(0.5+d_{\mathrm{i}}+d_{\mathrm{c}})))\bigg]\bigg\},
\end{split}
\end{equation}
in the perfect conductor, and 
\begin{equation}
\begin{split}
    E_{x} &= [1-\exp(-t (c/v_{\mathrm{A}})^2)] E_{\mathrm{v}0} \\
    &+ \left[\exp(-c_0 t) - \exp(-t (c/v_{\mathrm{A}})^2)\right] \bigg\{\left[-E_{\mathrm{v}0} \cos(\sqrt{\mathrm{S}_{\mathrm{c}} c_0}d_{\mathrm{c}}) + \sqrt{\frac{c_0}{\mathrm{S}_{\mathrm{c}}}} B_{\mathrm{v}0} \sin(\sqrt{\mathrm{S}_{\mathrm{c}} c_0}d_{\mathrm{c}})  \right]\\
    &- c_0 \left(\lvert z \rvert - 0.5 - d_{\mathrm{i}}\right)\bigg[\sqrt{\frac{\mathrm{S}_{\mathrm{c}}}{c_0}} E_{\mathrm{v}0} \sin(\sqrt{\mathrm{S}_{\mathrm{c}} c_0}\, d_{\mathrm{c}}) + B_{\mathrm{v}0} \cos(\sqrt{\mathrm{S}_{\mathrm{c}} c_0}\, d_{\mathrm{c}}) \bigg] \bigg\},
\end{split}
\end{equation}
\begin{equation}
\begin{split}
    B_y &= [1-\exp(-t (c/v_{\mathrm{A}})^2)]\sign(z)(\mathrm{S}_{\mathrm{c}} E_{\mathrm{v}0}\, d_{\mathrm{c}} + B_{\mathrm{v}0}) \\
    & - \sign(z) \left[\exp(-c_0 t) - \exp(-t (c/v_{\mathrm{A}})^2)\right]\bigg[\sqrt{\frac{\mathrm{S}_{\mathrm{c}}}{c_0}} E_{\mathrm{v}0} \sin(\sqrt{\mathrm{S}_{\mathrm{c}} c_0}\, d_{\mathrm{c}}) + B_{\mathrm{v}0} \cos(\sqrt{\mathrm{S}_{\mathrm{c}} c_0}\, d_{\mathrm{c}}) \bigg].
\end{split}
\end{equation}
in the insulator. This shows that including the displacement current effects and solving the complete Maxwell's equations again yields a solution that satisfies causality --- $E_x = B_y = 0$ at $t=0$ inside the end caps. Moreover, a few $(v_{\mathrm{A}}/c)^2$ time periods after $t=0$, displacement current effects become negligible and these solutions become identical to~\eqref{eqn:imperfect-conductor-E_x},~\eqref{eqn:imperfect-conductor-B_y},~\eqref{eqn:E_y_insulator}, and~\eqref{eqn:B_y_insulator}. Since the short-time solution is only dominant for $t \sim (v_{\mathrm{A}}/c)^2$ around $t=0$, it does not affect the long-time dynamics, the steady-state solution, or any of the results in this paper.

\section{Analytical solution of MHD equations inside the plasma in the presence of boundary layers}
\label{app:Analytical_solution}
In this appendix, we demonstrate how the one-dimensional equations~\eqref{eqn:ideal-MHD-momentum1}-\eqref{eqn:ideal-MHD-induction2}, which govern the plasma dynamics, can be further simplified and solved analytically if we make a few additional assumptions. We then compare the analytical solutions comprising the $u_y$ and $B_y$ profiles with the exact numerical solutions.

\begin{figure}
    \centering
    \begin{tabular}[b]{c} 
        \includegraphics[width=0.42\linewidth, trim = 0 0mm 0 0mm, clip]{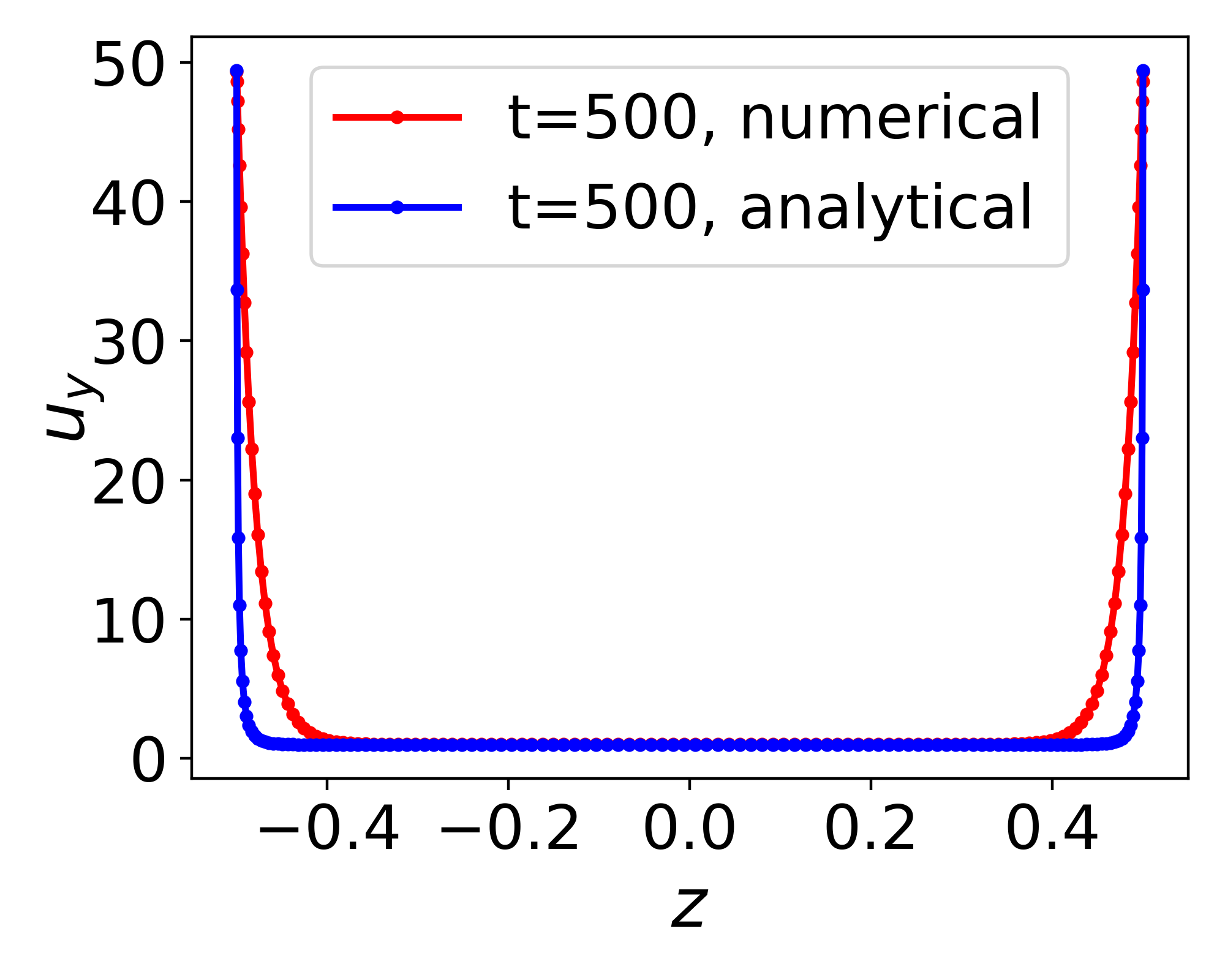} \\
        {\textit{(a)} $\mathrm{Ha} = 50$}
    \end{tabular} 
  \begin{tabular}[b]{c}
        \includegraphics[width=0.42\linewidth, trim = 0 0mm 0 0mm, clip]{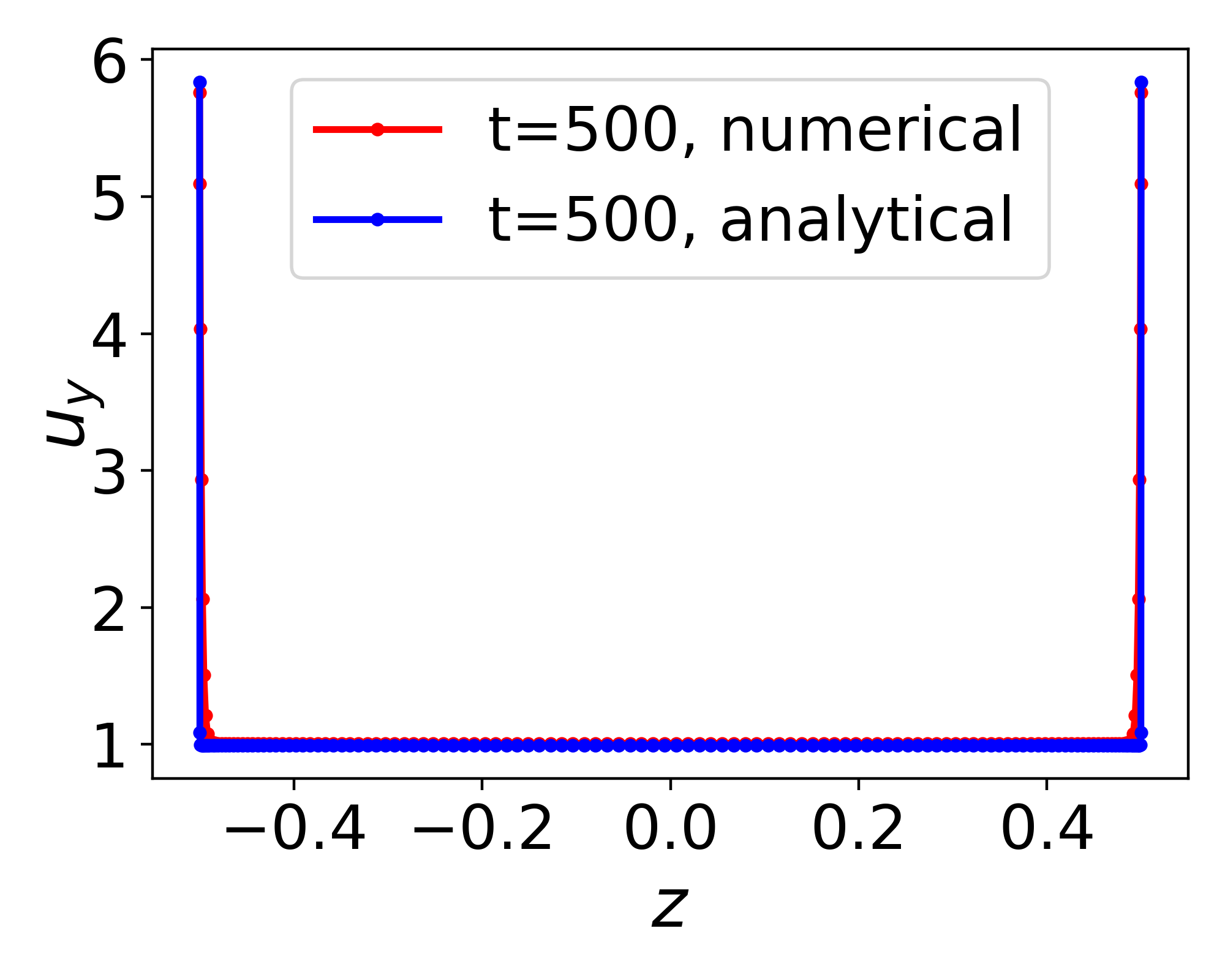} \\
        {\textit{(b)} $\mathrm{Ha} = 500$}
  \end{tabular}
    \caption{In this figure, we present a comparison between the analytical and numerical solutions close to steady state at two different Hartmann number values. The boundary layer approximation improves the agreement between analytical and numerical solutions. Note that plasma-generated flows have been scaled by $1/\epsilon$ due to their small size compared to the Alfv\'{e}n speed.}  
    \label{fig:uy_comparison}
\end{figure}

First, we decouple the equations~\eqref{eqn:ideal-MHD-momentum1}-\eqref{eqn:ideal-MHD-induction2} governing the plasma to obtain the following system of equations,
\begin{equation}
    \partial^2_t u_y = \partial^2_z u_y + \left(\frac{1}{\rm{Re}} + \frac{1}{\rm{Rm}}\right)\partial_t \partial^2_z u_y - \frac{1}{\rm{Ha}^2} \partial^4_z u_y + F_0 c_0 \exp(-c_0 t),
    \label{eqn:decouple-mom-1}
\end{equation}
\begin{equation}
    \partial^2_t B_y = \partial^2_z B_y + \left(\frac{1}{\rm{Re}} + \frac{1}{\rm{Rm}}\right)\partial_t \partial^2_z B_y - \frac{1}{\rm{Ha}^2} \partial^4_z B_y.
    \label{eqn:decouple-ind-1}
\end{equation}
Next, we utilize the fact that the dimensionless numbers $\mathrm{Re}$ and $\mathrm{Rm}$ are large for a typical fusion plasma and introduce a new small length scale, the Hartmann boundary-layer width corresponding to the Hartmann number $\mathrm{Ha} \equiv \sqrt{\mathrm{Re}\mathrm{Rm}}$. Mathematically, this corresponds to the introduction of an auxiliary small parameter 
\begin{equation}
    \delta \sim \frac{1}{\mathrm{Re}} \sim \frac{1}{\mathrm{Rm}} \sim \frac{1}{\mathrm{Ha}}, \delta \gg \epsilon, 
\end{equation}
where $\epsilon$ is the small parameter of the system used in Section~\ref{sec:section-2}. Introducing the small parameter $\delta$ allows us to separate our solution into a core solution described by the ideal MHD equations and a boundary layer solution determined by non-ideal effects. Due to the size of the dimensionless constants, we can solve the model in two regions: a large core region governed by
\begin{equation}
    \partial^2_t u_y = \partial^2_z u_y + F_0 c_0 \exp(-c_0 t),
\end{equation}
and a thin boundary layer near the domain walls governed by the equation
\begin{equation}
    \partial^2_z u_y = \frac{1}{\rm{Ha}^2} \partial^4_z u_y.
\end{equation}
These equations can be solved separately, and the different solutions can be combined to obtain an overall time-dependent solution. To further simplify the model, we impose a parity on the solutions. A general solution is considered admissible if it satisfies the parity conditions~\eqref{eqn:parity-conditions}, which allows us to solve equations only in one-half of the domain along the $z$-axis. We also limit this study to solutions that have a nongrowing time-dependent part, as we argue based on the findings of~\citet{hassam1999velocity, huang2001velocity} that turbulence is suppressed toward the edge due to the presence of a large velocity shear. Using these conditions, we solve~\eqref{eqn:decouple-mom-1}, and~\eqref{eqn:decouple-ind-1} for various quantities inside the plasma
\begin{equation}
    u_y = A_1 - A_2 \cosh(\mathrm{Ha}\, z) + \exp(-c_0 t) \left(\frac{F_0}{c_0} +  B_1 \cosh(-c_0 z) - B_2 \cosh(\mathrm{Ha}\, z) \right) ,
\end{equation}
\begin{equation}
    B_y = -F_0 z + A_2 \frac{\mathrm{Ha}}{\mathrm{Re}} \sinh(\mathrm{Ha}\, z)  + \exp(-c_0 t) \left(- B_1 \sinh(-c_0 z) + B_2\frac{\mathrm{Ha}}{\mathrm{Re}} \sinh(\mathrm{Ha}\, z) \right).
\end{equation}
Using~\eqref{eqn:Ohms-law-1}, we can write the lowest-order electric field
\begin{equation}
    E_x = -A_1 - \exp(-c_0 t)\left(\frac{F_0}{c_0} + B_1 \cosh(-c_0 z)\right),
\end{equation}
Note that $c_0 \ll \mathrm{Ha}$ for a consistent solution.
Finally, we ensure the consistency of the tangential electric and magnetic fields by matching the field inside the plasma with the values of fields inside the insulator, obtained by evaluating~\eqref{eqn:B_y_insulator} and~\eqref{eqn:E_y_insulator} on the boundary. This gives us the expressions for the coefficients $A_1, A_2, B_1$, and $B_2$ 
\begin{equation}
    A_1 = -E_{\mathrm{v}0},
\end{equation}
\begin{equation}
    A_2 = \frac{\mathrm{Re}}{\mathrm{Ha}} \frac{1}{\sinh(\mathrm{Ha}/2)} \left[\mathrm{S}_{\mathrm{c}} E_{\mathrm{v0}} d_c + B_{\mathrm{v0}} + \frac{F_0}{2}\right],
\end{equation}
\begin{equation}
\begin{split}
    B_1 = \frac{1}{\cosh(-0.5 \, c_0)}&\Bigg[(E_{\mathrm{v}0} - c_0 d_{\mathrm{i}} B_{\mathrm{v}0}) \cos(\sqrt{\mathrm{S}_{\mathrm{c}}c_0}d_{\mathrm{c}}) \\
    &- \sqrt{\frac{c_0}{\mathrm{S}_{\mathrm{c}}}} \left( B_{\mathrm{v}0} + \mathrm{S}_{\mathrm{c}} d_{\mathrm{i}} E_{\mathrm{v}0}\right) \sin(\sqrt{\mathrm{S}_{\mathrm{c}}c_0}d_{\mathrm{c}}) - \frac{F_0}{c_0}\Bigg],
\end{split}
\end{equation}
\begin{equation}
    B_2 = \frac{1}{\sinh(\mathrm{Ha}/2)} \frac{\mathrm{Re}}{\mathrm{Ha}} \Bigg\{\sinh(-c_0/2)B_1 - \left[\sqrt{\frac{\mathrm{S}_{\mathrm{c}}}{c_0}} E_{\mathrm{v}0} \sin(\sqrt{\mathrm{S}_{\mathrm{c}} c_0}\, d_\mathrm{c}) + B_{\mathrm{v}0} \cos(\sqrt{\mathrm{S}_{\mathrm{c}} c_0}\, d_{\mathrm{c}}) \right] \Bigg\}.
\end{equation}
Note that this model is only valid in the presence of a thin boundary layer. To better understand the boundary layer approximation, we compare these analytical solutions with the numerical solutions used in Figure~\ref{fig:uy_and_By_z} and present them in Figure~\ref{fig:uy_comparison}.

The comparison illustrates how well the analytical solutions perform close to steady-state conditions. Nevertheless, when far from steady state or in scenarios governed by a lower Hartmann number, the analytical model exhibits a significant deviation from the actual numerical solution. Hence, we used numerical solutions for all the analyses presented in this study.

\bibliographystyle{jpp}
\bibliography{jpp-instructions}
\end{document}